\documentclass[preprint,showpacs,preprintnumbers,amsmath,amssymb]{revtex4}

\usepackage{dcolumn}
\usepackage{bm}
\usepackage{epsfig}

\begin{document}

\preprint{APS/123-QED}

\title{On the Mechanism of Townsend Avalanche for Negative Molecular Ions}

\author{M.P. Dion}
 \altaffiliation{email: mpdion@umd.edu}
\author{C.J. Martoff}%
 \altaffiliation{Corresponding author: jeff.martoff@temple.edu}
\author{M. Hosack}
 \altaffiliation{email: hosack@fnal.gov}
\affiliation{%
Temple University\\
Philadelphia, PA 19122
}%

\date{\today}

\begin{abstract}
Time projection chambers drifting negative ions (NITPC) instead of
electrons have several advantages. \ A NITPC can operate at very high
reduced drift fields without diffusion runaway, and the readout
digitization sampling rate requirement is considerably relaxed due to
the low drift speed of negative ions. \ The initiation of Townsend
avalanches to allow gas gain in these devices has not been understood
until now. \ It is shown here that the avalanche in low pressure
CS$_2$ vapor is most likely initiated by collisional detachment of
the electron from the negative molecular ion. \ In mixtures of Nitromethane
vapor with CO$_2$ the mechanism appears to be more complex.
\end{abstract}

\maketitle

\section{\label{intro}Introduction\protect\\}
The Time Projection Chamber (TPC) is one of the most productive and widely used
detector types in nuclear and particle physics. \ Recently, TPC's have been applied in experiments such as dark matter searches\cite{drift,NEWAGE,miuchi} and X-ray polarimetry\cite{black07a} where short tracks of very low energy particles must be measured. \ In these situations the detectors are often operated at pressures of a fraction of an atmosphere to extend the track lengths. \ However, with these applications suppression of both transverse and longitudinal diffusion is also of paramount importance. \ In high energy physics TPC's, the transverse diffusion is usually suppressed by applying a magnetic field along the drift direction, but this has no effect on the longitudinal diffusion and it is impractical and costly for many specific applications.

The Negative Ion TPC (NITPC) was invented to achieve the lowest possible track length thresholds\cite{orignitpc}. \ Negative ions can be drifted long distances at very high reduced drift fields (and zero magnetic field) with diffusion at the thermal lower limit.  Under these conditions the best ``cool" electron-drift gases would suffer from runaway diffusion due to heating of the electrons by the drift field\cite{saulibible}.  Thermal limit diffusion over a drift distance of 50 cm at $>$15 $\frac{V}{cm\cdot Torr}$ has been achieved in routine operation\cite{drift2}.

As well as low diffusion, NITPC's must produce enough gas gain to give usable signals. \ Negative ions have been found to produce gas gains greater than 1000 with proportional wires\cite{drift}, GEMs\cite{gemni} and bulk micromegas\cite{bmmg}. In electron gases, gas gain results when some electrons in the very high electric fields of gain structures attain energies comparable
with the ionization energy of the fill gas. \ Relatively moderate electric
fields (of order 100 V/(cm Torr) are sufficient to initiate avalanche gain in
electron gases because the light electrons are poorly thermally coupled to the heavy gas atoms, even in ``cool gases" like CO$_2$ which have substantial inelastic scattering cross sections.

Negative ion gases generally show lower gain and require higher operating 
voltages than electron gases\cite{gemni}. \ Although the electron affinity of negative ion capture agents is much lower than the ionization energy of gases, the drift field needed to free electrons from ions is almost certainly higher 
than that needed to initiate an avalanche in an electron gas.  Once electrons are freed from the ions in such high fields an avalanche should develop immediately.  However, the mechanism of producing the initial free electrons in negative ion avalanches has not been explored until now. \ Understanding this mechanism will allow further development of negative ion mixtures to be tailored to specific applications.  

Here we study the gas gain and effective drift mean free path as a function of applied drift fields in a single-wire proportional counter with attached drift space. \ This data allows us to clarify the mechanism of gain initiation and production in the negative ion gases carbon disulfide (CS$_2$)\cite{orignitpc} and in Nitromethane:carbon dioxide (CH$_3$NO$_2$:CO$_2$)\cite{nitro} mixtures. \ The data was analyzed using the Diethorn formalism\cite{diethorn} and compared to a simple model based on the ion collision energy needed to free electrons given the  binding energy (electron affinity) of the anions.

\section{Apparatus}
Negative ion drift velocity and gas gain were measured at low to moderate drift fields using a single{}-wire proportional counter coupled to a homogeneous{}-field drift region. \ The drift stack was eight drift field plates wired with a voltage divider chain of high voltage resistors in order to create a linear graded potential from the cathode to the vicinity of a 15 $\mu$m diameter gold plated tungsten wire. \ Drift fields up to 4.0 $\times$ 10$^{4}$ V/(m Torr) were obtained with this apparatus for drift velocity measurements.

Primary photoelectrons were generated at a tin photocathode attached to the drift{}-cathode. \ The photocathode was arranged so that it could be cleaned between runs using a glow discharge in pure argon. \ This was essential for maintaining the photoelectron yield. 

UV light flashes from an EG\&G Flash{}-Pak\cite{egg} were admitted into the stainless steel bell jar through a quartz window, passed through a hole in the proportional counter body, and struck the photocathode producing photoelectrons. \ The standard internal capacitors of the Flash{}-Pak were augmented with additional HV capacitors to give a stored energy of about 0.2 Joule per pulse. \ The Flash{}-Pak was triggered by an external pulser, from which a time{}-zero signal was derived on the oscilloscope. \ Charge pulses from the wire were sent through an EG\&G Ortec 142PC charge sensitive preamplifier and shaped by an EG\&G Ortec 572 amplifier. \ Data was acquired from the digital oscilloscope traces transferred to a desktop PC via a National Instruments GPIB-to-ethernet transceiver\cite{nigpib}.

Between runs the bell jar was repeatedly pumped out to $\sim$ 20 mTorr and backfilled with the next gas mixture to be tested. \ Pumping was through a removable LN-cooled trap and a MicroMaze trap, preventing the toxic and hazardous vapors from entering the mechanical pump oil and exhaust. \ A simple gas manifold allowed for the addition of gas mixtures, including the organic vapors. 

Drift speeds were measured in this setup using the time difference between the light pulser trigger and the wire signal. \ A GARFIELD\cite{magboltz} simulation of the apparatus was used to separate the drift speed in the constant field drift region from that in the cylindrical region near the wire.  

Absolute gas gains were measured using a slightly modified apparatus. A 1 $\mu$Ci $^{241}$Am source, a collimator with solenoid-operated shutter, and a Si trigger detector were mounted behind a slit which replaced the field cage plate nearest the drift cathode. \ The Si detector gave a pulse indicating that an $\alpha$-particle had traversed the drift space, and the slit defined the track length from which charge was collected and drifted to the amplification region. \ SRIM\cite{srim} energy loss tables were computed and used
with an estimated ionization yield parameter "W" of 20 eV/ionization electron for all gas mixtures, to determine the number of primary electrons collected. \ The electronic gain was known from manufacturer's data and checked with pulser measurements, thus allowing the absolute gas gain to be computed.

\section{Results and Discussion}
\label{mobility}
\subsection{Drift Speed and Mobility}
Carbon Disulfide (CS$_2$) and Nitromethane (CH$_3$NO$_2$) are known to work as   electronegative capture agents\cite{orignitpc,nitro} in gas detectors. \ CS$_2$ is usable as the pure vapor, while carbon dioxide (CO$_2$) is added to Nitromethane to allow efficient electron capture\cite{nitro,saulibible}.

\begin{figure}[htp]
\begin{center}\thicklines
\epsfig{file=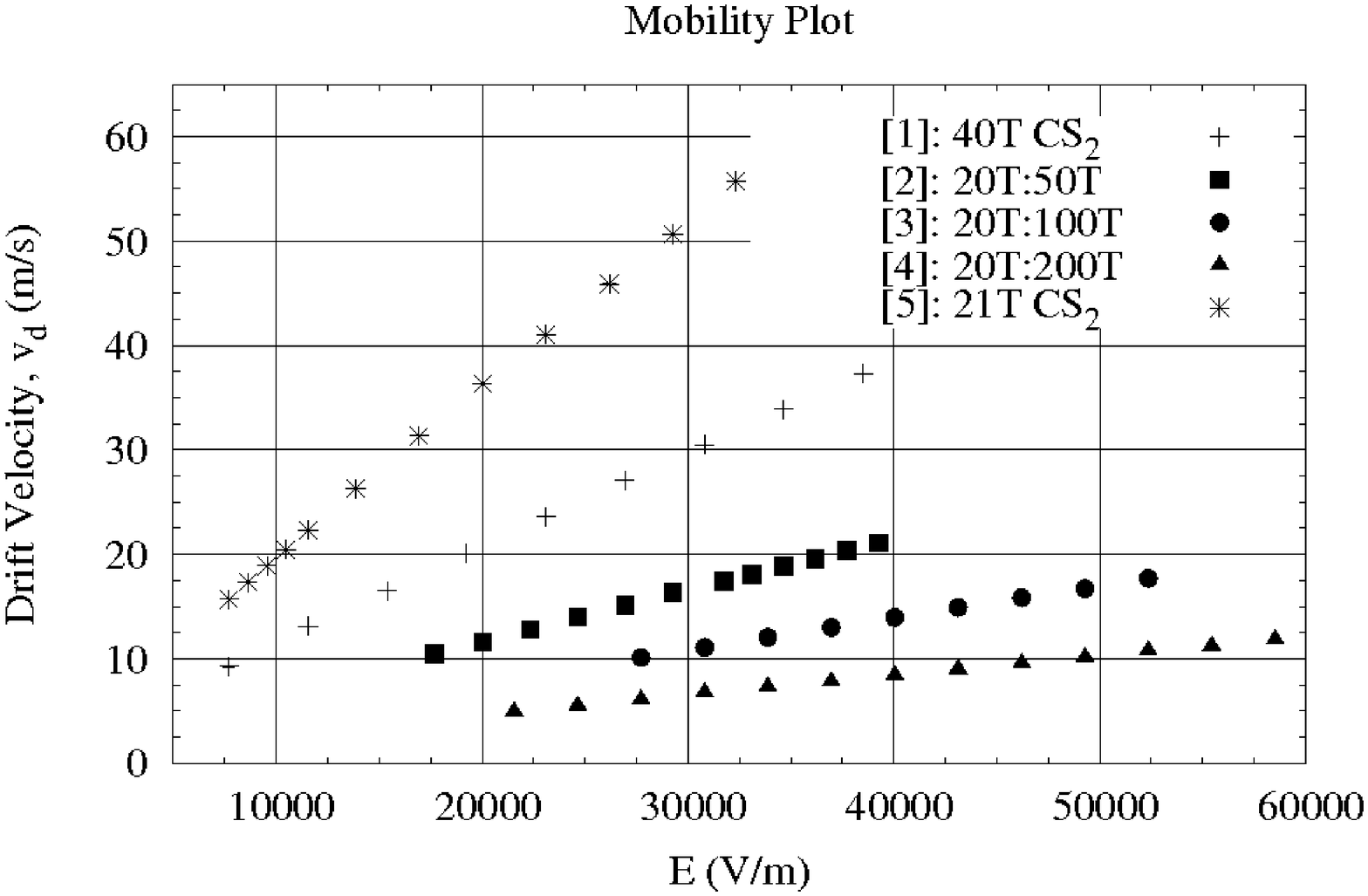,height=3.0in}
\caption{Drift speed versus drift field for the negative ion gas mixtures studied.  Mixtures [2] - [4] in legend correspond to CH$_3$NO$_2$:CO$_2$ mixtures.}
\label{velall}
\end{center}
\end{figure}

Results for drift speeds as a function of electric field in the drift region are shown in Figure \ref{velall}, for CS$_2$ at 21 and 40 Torr and for Nitromethane:CO$_2$ mixtures of 20:50, 20:100, and 20:200 Torr. \ There is a linear relationship between the drift velocity ${v_d}$ and the electric field $E$ in Figure \ref{velall} which indicates that in the range of drift fields measured, the ions remain thermally coupled during drift, leading to a drift velocity which is proportional to the drift field $E$, and hence a constant mobility $\mu$. \ In Table \ref{moball} the fitted slopes of the drift speed vs. applied drift field curves are tabulated.

\begin{table}
\centering
\begin{tabular*}{0.55\textwidth}%
     {@{\extracolsep{\fill}}lc}
\hline
Gas mixture & $\mu$  \\
\vspace{0.2in}
& ($\frac{m^2}{V \cdot s}$)  
\\\hline

21T CS$_2$    & 1.6$\times$10$^{-3}$ $\pm$ 8.6$\times$10$^{-6}$\\

40T CS$_2$    & 9.0$\times$10$^{-4}$ $\pm$ 4.2$\times$10$^{-6}$ \\

20T:50T CH$_3$NO$_2$:CO$_2$   & 4.6$\times$10$^{-4}$ $\pm$ 2.9$\times$10$^{-6}$ \\

20T:100T CH$_3$NO$_2$:CO$_2$  & 2.9$\times$10$^{-4}$ $\pm$ 1.1$\times$10$^{-6}$ \\

20T:200T CH$_3$NO$_2$:CO$_2$  & 1.9$\times$10$^{-4}$ $\pm$ 1.5$\times$10$^{-6}$\\\hline

\end{tabular*}
\caption{Mobility of negative ion gas mixtures studied.}
\label{moball}
\end{table}

The slope of drift speed vs. applied drift field for CS$_2$ is seen to be very nearly twice as large at 21 Torr as at 40 Torr. \ The corresponding mobility $\mu = v_d / (E/P)$ is therefore approximately constant for CS$_2$ independent of field or pressure. \ This is the expected behavior for the low field limit of drift in either electron or negative ion gases, with constant (thermal) energy of the charge carriers and hence constant collision mean free path\cite{rolbloom}.  

The drift speeds for the Nitromethane:CO$_2$ mixtures are linear with respect to the drift field (Figure \ref{velall}). However, while the slopes of drift speed vs. applied drift field have the expected decreasing trend with increasing pressure, they do not scale linearly with the total pressure of the mixture. \ In the low field limit, the drift speed $v_d$ is determined by an effective mean free path $\lambda$ according to $v_d = \frac{eE}{m}\lambda / v_{th} f(m,M)$ where $f(m,M)$ is a momentum transfer efficiency factor that depends on the masses of the charge carrier and the gas molecules. \ In this picture the effective mean free path in the Nitromethane mixtures must be changing not only with total pressure but with the composition of the gas.

\subsection{Gas Gain}

Gas gain measurements were analyzed using a Diethorn plot\cite{diethorn,egas,edieth,diele}. \ This plot is based on an analytic calculation of the gain in a cylindrical proportional counter as a function of applied voltage, under the assumptions that the electric field throughout the gain region is proportional to 1/r, and that the First Townsend Coefficient\cite{rolbloom} increases linearly with electric field. \ The first of these assumptions is well satisfied for most arrangements involving gain near a wire, including the present one. \ The second assumption is more questionable, but experience shows that it is sufficiently accurate to give informative results when applied in the restricted range of field values important for gain around a proportional wire. \ The resulting expression relating gain to applied voltage is cast in terms of two parameters: E$_{min}$, the electric field where the avalanche starts (referred to as the ``starting field"), and $\Delta V$, the potential difference through which the avalanche charge doubles:

\begin{equation}
\frac{ln (G) \ ln(\frac{b}{a})}{V} = \frac{ln 2}{\Delta V} \cdot ln \left(\frac{V}{ln(\frac{b}{a})a E_{min}} \right)
\end{equation}

Here a and b are the radii of the proportional wire and the cylindrical cathode respectively.

This development predicts a straight-line plot for ln(G)/V vs. V with G the gas gain and V the applied voltage. \ The slope and intercept of the linear fit determine  $\Delta$ V and   E$_{min}$ respectively. \ These two ``Diethorn parameters" reflect the physics of avalanche development in a particular gas mixture, and have been studied for many electron gas mixtures\cite{egas,edieth}. \ To our knowledge this is the first Diethorn analysis applied to negative ion gases.

\begin{figure}[htp]
\begin{center}\thicklines
\epsfig{file=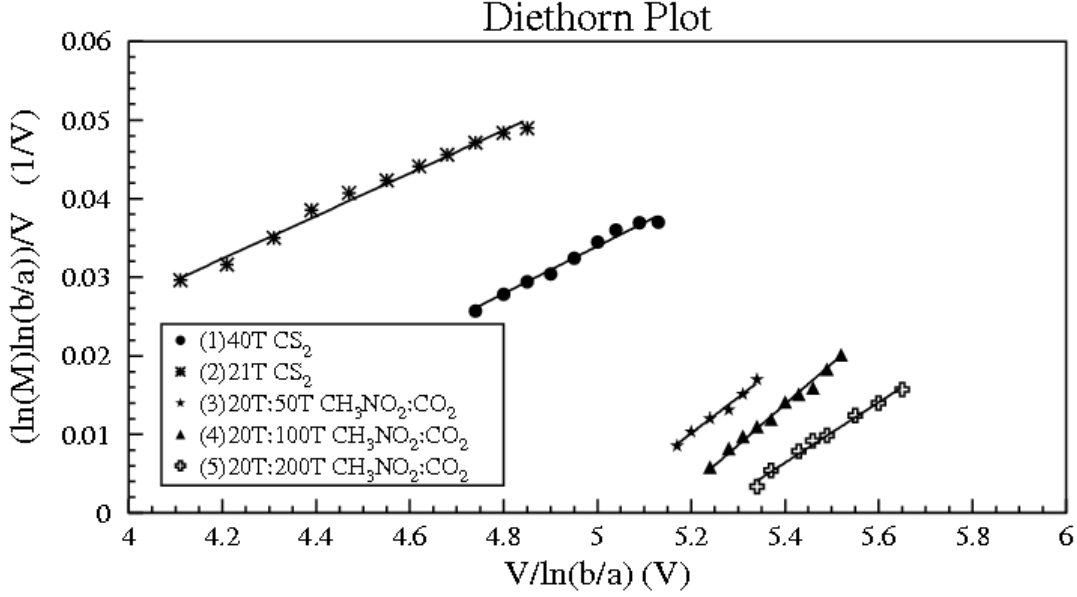,height=3.2in}
\caption{Diethorn plots for several negative ion gas mixtures studied in the present work.}
\label{diethorn}
\end{center}
\end{figure}

Figure \ref{diethorn} shows Diethorn plots for the negative ion gas mixtures studied. \ Table \ref{parameter} contains the Diethorn constants obtained by fitting the linear portions of the curves. For comparison, the Diethorn constants for 90\%:10\% argon:methane\cite{egas} (P10) are shown as (6) in Table \ref{parameter}.

\begin{table}[htp]
\centering 
\begin{tabular*}{0.7\textwidth}%
     {@{\extracolsep{\fill}}lccc}
\hline
Data Set & $\Delta$V  & E$_{min}$  & E$_{min}$/P$_{total}$\\
& (V) & ($\frac{V}{m}$) \\
\\\hline

Gas (1); 21T CS$_2$ & 21.5 & 3.34$\times$10$^6$ & 1.59\\

Gas (2); 40T CS$_2$ & 21.5 & 6.94$\times$10$^6$ & 1.74\\

Gas (3); 20T:50T & 10.8 & 20.8$\times$10$^6$ & 2.97\\

Gas (4); 20T:100T & 13.8 & 22.5$\times$10$^6$ & 1.88\\

Gas (5); 20T:200T & 17.8 & 25.1$\times$10$^6$ & 1.14\\

Gas (6); P10 & 23.6 & 4.8$\times$10$^6$ & 0.063\\

\hline

\end{tabular*}
\caption{Diethorn parameters E$_{min}$ and $\Delta V$ for the negative ion gas mixtures studied.  Gases (3) - (5) are CH$_3$NO$_2$:CO$_2$ mixtures.}
\label{parameter}
\end{table}

The data shows that $\Delta V$ for negative ion gases is similar or even smaller in magnitude when compared to P10. \ However, $E_{min}$ is much larger for the negative ion gas mixtures than for the electron gas. \ With the exception of 21T CS$_2$ all negative ion gas mixtures have starting voltages much higher than P10. \ The difference between electron and negative ion gases is much more striking when expressed as (pressure reduced) starting voltages $E_{min}/P$, as shown in Table \ref{parameter}.

\subsubsection{Carbon Disulfide (CS$_2$)}
The Diethorn parameter E$_{min}$ for CS$_2$ is seen to scale with pressure within $\sim$ 5\%. \ This is the classic behavior expected for collision-controlled processes in gases\cite{saulibible}. \ For example, the mobility is a constant independent of pressure for a given electron gas at low drift fields. \ This scaling results from the 1/P dependence of the collision mean free path ($\lambda$) in the gas--E/P scaling is actually E$\cdot \lambda$ scaling, and this quantity is proportional to the mean velocity $\frac{eE}{m}\frac{\lambda}{v_{th}}$ gained between collisions by thermal carriers from the drift field. \ This velocity increment determines the drift speed.

Like the mobility, E$_{min}$ in electron gases also scales with pressure\cite{diele}. \ This scaling suggests that we should examine the ``reduced starting field" E$_{min}$/P$_{total}$, as shown tabulated in the last column of Table \ref{parameter}. \ As previously mentioned, this reveals E$_{min}$/P values are orders of magnitude larger for CS$_2$ than for the electron gas P10. \ Such a large difference in E$_{min}$/P$_{total}$ can be understood by considering the energetics of avalanche initiation.

Avalanches in electron gases begin when the applied field drives the electrons' energy distribution high enough to give a finite probability of secondary ionization by electrons in the high energy tails of the distribution\cite{mcdanielcoll}. \ Even for well-quenched electron gases, the energy distribution is distinctly non-thermal above a few V/(cm Torr), due to the inefficient energy exchange in electron-molecule collisions. 

In contrast, negative ion gases exchange energy in ion-molecule collisions much more efficiently. \ The mobility remains constant and the free-flight diffusion\cite{rolbloom}  scales with 1/$\sqrt{E}$ up to much higher fields\cite{orignitpc,nitro}, indicating that the drifting ions remain in thermal equilibrium with the gas. \ To initiate an avalanche, electrons must first be liberated from the slow ions. \ If this occurs in a region of a sufficiently high field, the electrons will immediately gain enough energy from the field to evade capture (the electron capture cross sections of CS$_2$ and Nitromethane are large only for thermal electrons\cite{chen}). \ Without losses to capture, these free electrons rapidly become sufficiently energetic to cause traditional Townsend avalanche multiplication. \ Apparently the limiting step is the liberation of electrons from the ions.

The values of $\Delta V$ are similar for all the gases in Table \ref{parameter} and the magnitudes are all similar to the ionization potentials of the gas mixtures. \ This is at least consistent with the above picture of free electron avalanches initiated by collisional detachment of electrons from negative ions.

Collisional neutralization of CS$_2$ molecular ions in thermal equilibrium with CS$_2$ vapor could occur by inelastic ion-molecule collisions if the CM energy exceeded the 0.6 eV electron affinity\cite{chen} of CS$_2$. \ If the drifting ions gain no energy on average from
the drift field, such collisions could only occur if the field were so strong that an ion could gain kinetic energy from the field equal to twice the electron affinity, in one mean free path. \ (The factor two arises from the LAB$\rightarrow$CM transformation for equal masses). \ This argument suggests the following criterion for collisional release of electrons from anions:
\begin{equation}
E_{min} \lambda \doteq 2*EA ,
\label{theeqn}
\end{equation}
where $EA$ is the electron affinity. 

To check Equation \eqref{theeqn}, the effective mean free path $\lambda$ can be estimated from the measured drift speed in CS$_2$.  
Using\cite{mcdanieldiff}:
\[
\lambda = \frac{v_d \sqrt{3MkT}}{eE}
\]    
where T=300 K and M is the mass of the CS$_2$ molecule. \ One obtains for 21 Torr (40 Torr) CS$_2$, $\lambda$ = 0.42 (0.23) $\mu$m, corresponding to an effective collision cross section of 3.5 (3.4) $\times$10$^{-18}$m$^2$ respectively. \ (This value compares reasonably to the elastic {\em electron}-ion cross section for CS$_2$ of 10$^{-18}$ m$^2$ below 0.5 eV electron energy used in the MAGBOLTZ program\cite{magboltz}.) \ Substituting this mean free path into Equation \eqref{theeqn} predicts for 21 (40) Torr CS$_2$ the values E$_{min}$ = 2.6 (4.9)$\times$10$^6$ V/m.  These estimates are in startlingly good (20\%) agreement with the values in Table \ref{parameter}, considering the oversimplified model being used.

The approximate constancy of E$_{min}$/P for pure CS$_2$ and the semi-quantitative agreement of the collisional estimate given above with the measured E$_{min}$ values, supports the conclusion that avalanches in CS$_2$ are initiated by collisional detachment of electrons from ions, followed by a normal Townsend avalanche of the free electrons.    

\subsubsection{Nitromethane:CO$_2$ Mixtures}

In contrast to CS$_2$, the E$_{min}$ values for the Nitromethane:CO$_2$ 
mixtures in Table \ref{parameter} are even larger. \ In fact, the field at the surface of the wire in the gain measurements for these mixtures ranged from 23-36 MV/m, not
much larger than the fitted  E$_{min}$ values. \ Nor do these values scale with pressure. \ In the context of the model presented above for CS$_2$ avalanches, the non-scaling could be explained if the ion-molecule collision cross sections for Nitromethane and CO$_2$ were very different. \ The mean collision cross section and hence the mean free path would not scale in a simple way with pressure as the gas composition changed.  

\begin{table}[htp]
\begin{center} 
\begin{tabular*}{0.85\textwidth}%
     {@{\extracolsep{\fill}}lcccc}
\hline
Data Set & E$_{min}$ & $\lambda$  & E$_{min} \cdot \lambda$ & $\frac{M}{\mu} \cdot$EA\\
& (MV/m) & ($\mu$m) & (eV) & (eV)\\\hline

21T CS$_2$ &  3.34 & 0.42 & 1.4 & 1.2\\

40T CS$_2$ &  6.94 & 0.23 & 1.56 & 1.2\\

20T:50T CH$_3$NO$_2$:CO$_2$ & 20.8 & 0.092 & 1.91 & 2.6\\

20T:100T CH$_3$NO$_2$:CO$_2$ & 22.5 & 0.056 & 1.26 & 2.6\\

20T:200T CH$_3$NO$_2$:CO$_2$ & 25.1 & 0.036 & 0.91 & 2.6\\

P10 & 4.8 & 0.19\cite{p10drift} &- &- \\

\hline
\end{tabular*}

\caption{Measured E$_{min}$ values (V/m), effective mean free paths $\lambda$ ($\mu$m) estimated from drift speed data, and energy gain per mean free path E$_{min} \cdot \lambda$ (eV).  The reduced mass for ion-majority molecule collisions is $\mu$ and the ion mass is M.}
\label{mixtures}
\end{center}
\end{table}

However, the E$_{min}$ values for these mixtures do not scale inversely as the actual mean free paths estimated from the drift speeds, either. \ Table \ref{mixtures} illustrates this. \ Multiplying E$_{min}$ by $\lambda$ does give numbers of the same order as EA times the
CM$\rightarrow$ LAB mass factor (2.6 for Nitromethane on CO$_2$). \ However, these values monotonically decrease with increasing pressure, changing by a factor of two for a change of a factor of three in total pressure. \ The simplified model gives about the right magnitude
for E$_{min}$, but  fails to track the pressure dependence in these mixtures as both the composition and total pressure change.

The very high E$_{min}$ values for the mixtures suggest that some assumptions of the highly simplified treatment used above may be violated. \ The Diethorn analysis is based on the assumption of a linear increase of the first Townsend coefficient with electric field. \ The good linearity of the Diethorn plots suggests that this assumption is not too badly violated over the range of fields where avalanches take place, despite the fields being high enough to drive even negative ions significantly out of thermal equilibrium with the gas\cite{rolbloom}. This may account for the rather low and falling values of E$_{min}$$\cdot$$\lambda$ in Table \ref{mixtures}. \ At such high fields the collisional detachment is occurring from collisions of ions in the tails of an elevated thermal-like energy distribution. \ The shift of the energy 
distribution is somewhat nonlinear with increasing E.

Another effect which may complicate the avalanche initiation process in these mixtures  is the capture of energetic electrons by CO$_2$. \ Unlike CS$_2$ and Nitromethane, CO$_2$ captures non-thermal electrons with energies above 3.3 eV\cite{co2attach}. \ This may introduce a second mean free path, leading to a more complicated density dependence.

\section{Conclusions}
Diethorn analysis of proportional counter gas gain in pure CS$_2$ has been carried out. \ The values and pressure scaling of the avalanche ``starting fields" E$_{min}$ are consistent with avalanche initiation by collisional detachment of electrons from CS$_2$ molecular anions, followed by ordinary Townsend avalanche of the freed electrons. \ Processes such as field ionization would give a constant $E_{min}$ independent of pressure and are ruled out by the data.

A similar analysis of CH$_3$NO$_2$:CO$_2$ mixtures shows that the simplified collisional detachment model predicts values of E$_{min}$ of the same order as the experimental results. \ However the values do not scale with pressure as predicted by the model. \ It is suggested that the extremely high fields needed for detachment in these mixtures, along with the capture of energetic electrons by CO$_2$, may account for these effects. 

\bibliographystyle{unsrt}
\bibliography{neg_diethorn}

\end{document}